# Design of Locally E-management System for Technical Education Foundation- Erbil

**Assistant Prof. Dr. Ayad Ghany Ismaeel**  &  **High Diploma. Engineer Dina Y. Mikhail**

Department of Information Systems Engineering
Technical College- Foundation of Technical Education. Erbil-IRAQ
E-mail (dr_a_gh_i@yahoo.com)

## Abstract

Until now, there is no e-management and automation necessary for the operations/ procedures of the departments in the Technical Education Foundation Erbil, and the foundation like any other organization in Kurdistan region is not connected to the network, because there isn't infrastructure for that purpose. To solve this problem, comes the proposal "**D**esign of **L**ocally e-**M**anagement **S**ystem **for T**echnical **E**ducation **F**oundation- Erbil", which is called **DLMS4TEF**.

DLMS4TEF's requirements are divided into hardware and software, as hardware will need Fast-Ethernet (LAN) technology to connect the departments of the Foundation via Client/Server network later, when an infrastructure is established for e-governments/e-management, it may be extended to the campus network. The software is represented by installing windows server to implement the proposal design of DLMS4TEF, PHP script is used as web programming that supports the server, where as the HTML and JavaScript are used to support the client side. The dynamic DLMS4TEF will be based on relational database, which is created by using MySQL, to support processing hundreds of queries per second, and the Kurdish Unicode to support Kurdish fonts of GUI's, Moreover, for security DLMS4TEF allows each department in the Foundation to enter its own section and prevent accessing other sections by using HTAccessible program which allows the user to access by using his IP address and his computer only.

The important conclusions and advantages of applying DLMS4TEF are: making backup to DLMS4TEF's databases using the option (zipped) which allows them to reach the size of ~3% of the original database size, sufficient security techniques, through achieving levels of security, hidden access to the administrator section, and finally DLMS4TEF, when compared with the traditional methods and Oman's project, shows the same efficiency of some, if not better, features of Oman's project.

# 1- Introduction

## 1.1  Overview

One word it is hard, if not impossible, to find a field in which computers and internet cannot help, obviously computer system grows up quickly in our daily life, it can fix or solves our problem faster than any other systems and even more accurate, so it will be a good idea is to computerize any project.

Web application is a term referring to a group of new techniques and applications retina which lead to change the behavior of the internet. It allows to link people and data in new and more effective ways, at a lower cost, in ways that were never possible by using traditional software [1]. Web applications also provide an added layer of security by removing the need of the user to have access to the data and back end servers. Web application systems need only to be installed on the server placing minimal requirements on the end user workstation. This makes maintaining and updating of the system much simpler as usually done on the server. Any client updates can be deployed via web server with relatively easy. There are many web applications such as E-learning which means Electronic-learning, E-business, E-marketing, E-management etc [2]. The e-Management or E-management use of information technology to improve the management of government, from streamlining business processes to maintaining electronic records, to improve the flow and integration of information. The term of e-Management describes the applications that will arise from the intersection of Management and e-Science. In other words, e-Management is not e-Commerce; rather, it is a more specific vision of how management and associated processes can benefit from grid computing **[3].** As examples of e-Management/e-Government are Muscat, Dubai, and the project of Greater Oman Municipality, which is serve a large section of costumers **[4].**

Till now there is no e-Management to manage the departments of Foundation of Technical Education- Erbil as a web application which constrains you to leave your chair, uses a large number of workers, and waste of time, effort, and energy. To solve this problem in Kurdistan Region where there are no resources to build E-management, is to concentrate on the importance of the modern infrastructure which is similar to the fiber-optic backbone as a first stage, and later to reach to the completed e-management (i.e. become E-government later) upon the availability of its requirements, as a result the proposal "Design of Locally E-management System for Technical Education Foundation- Erbil" comes as a local website application which is called **DLMS4TEF**. It will be used for the completion of the services division from the traditional manual method into electronic format for the optimal use of time, money and effort, to eliminate the gap between the organizational management among officials in higher floors and workers in the bottom, also a less need for the use of paper and archives.

## 1-2 Aims of DLMS4TEF

The main aims of this research are design and implement a local website for an e-Management to link between information and technology for Foundation of Technical Education refers to it previously as DLMS4TEF, and to rely on the modern electronic devices, including the computer for the purpose of organization and management of the business as a first stage of completed E-government later. DLMS4TEF will facilitate processing the application for every contacting person by making him to deliver the application in the reception of the foundation and wait until it is finished in the same day or might come again in the next day, and DLMS4TEF will satisfy the following features:

1. Facilitates the process.
2. The pace of work completion and time saving.
3. Gives the highest accuracy in work product.
4. Measures to facilitate communication within Technical Education Foundation.
5. The less need of employees.
6. Reduces the reliance on human role through the transition of leadership that is based on tasks or workers to the leadership that is based on technology.
7. Reduces the cost, by providing optimal use of time and efforts.
8. More security.
9. Monitors and determines the weakness in the performance, i.e. satisfy high performance.
10. Minimizes losing in the transactions and determines the responsible.

## 2- The Proposal of DLMS4TEF

## 2-1 Configuration of DLMS4TEF

The configuration of DLMS4TEF is divided into two parts:
1. **Hardware configuration**: which is related with to the equipments in DLMS4TEF, the campus network is a building or a group of buildings all connected into one enterprise network that consists of many local area networks (**LANs).** The campus network topology is primarily LAN technology connecting all the end systems within the building. Campus networks generally use LAN technologies, such as Ethernet, Token Ring, Fiber Distributed

Data Interface **FDDI**, and Asynchronous Transfer Mode **ATM [5]**. To connect the Foundation's building, it has been suggested Ethernet LAN technology, because today **[6]:**

   i. It is the most popular type of network in the world;
   ii. It has its fixed packet size (64-1516 Bytes) for all type of Ethernet, so it makes upgrade from one type to another easy and with little cost;
   iii. It is easy to implement, easy to manage; and
   iv. The cost of ownership is relatively lower than other technologies.

When trying to connect more than one computer together, there are also many different ways to get the job done. Some ways are better than others in certain situations, and it is very beneficial to get started in the right direction when networking computers, not only because of usefulness, but also because of security issues. There are two types of network, peer to peer, and client server, for DLMS4TEF will suggest client/server network, to connect the entire foundation's departments.

Now, it is suggested one workstation for each Foundation's department, except for the administrator department, to connect clients with server through Fast Ethernet switch (100 MB/s) by using Unshielded Twisted Pair (UTP) as media to connect. Because DLMS4TEF must be reached in the future to LAN for each department in Foundation, to support multimedia and real time applications, these applications required that bandwidth can range anywhere from 100 KB/s to 70-100 MB/s **[6].**

2. **Software configuration:** that is related to the used software of DLMS4TEF, this section discusses software configurations that are dividing into two parts: the first is server software that is related or used in server, and the second is client software configurations that are related or used it in client.
  A. **Server Software:** Windows server because it is friendly and easy in use, WAMP5 Server version 1.7.4, Kurdish Unicode, HTAccess, modern version of Microsoft office, Adobe Reader, WinRAR, and AutoCAD.
  B. **Client Software:** the important of it are:
     i. Kurdish Unicode: clients have been needed to write/enter information in DLMS4TEF using Kurdish font, and so they will need to install Kurdish Unicode in each client.
     ii. Others software can support the documents such as Firefox/Microsoft Internet Explorer, modern versions of Microsoft office, WinRAR, AutoCAD, and Adobe reader.

## 2-2 DLMS4TEF's Map

To design DLMS4TEF website structure, there are two options *hierarchical* and *flat*, for DLMS4TEF a hierarchical construction has been selected, so the site map of DLMS4TEF consists of three levels: *Level1*(index/home page), which explains the structure of the Foundation and covers the functions by using hyperlinks; *Level2* contains the sections and departments of the Foundation; and *Level3* which takes care of the functions/tasks of each section and department in the foundation as shown in figure (1).

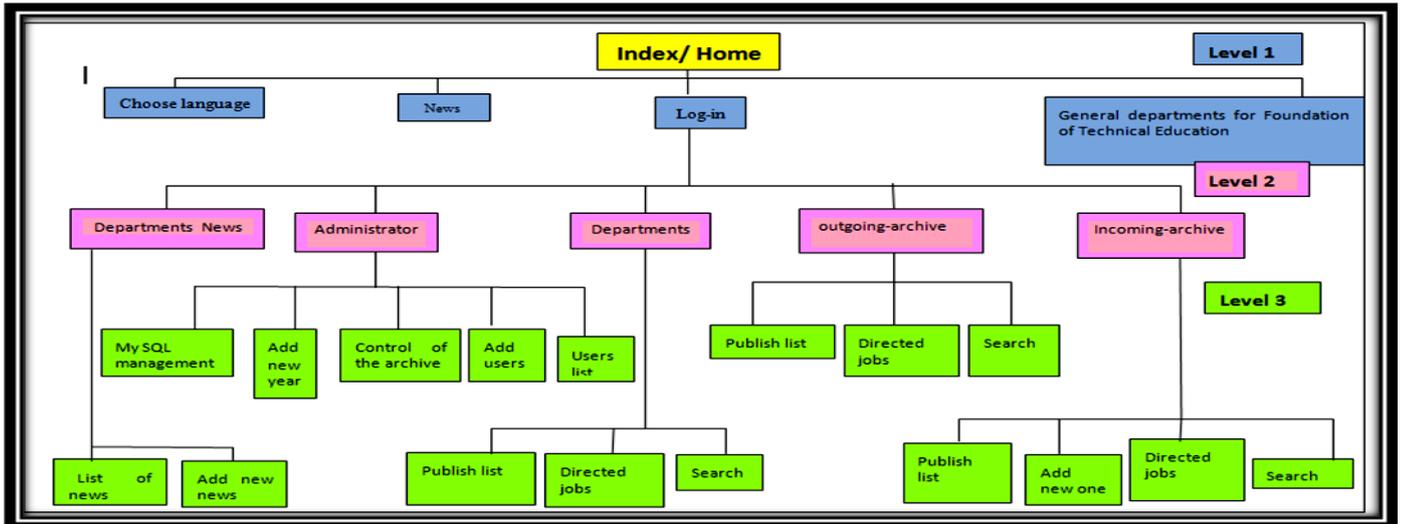

*Figure (1): DLMS4TEF's map*

## 2-3 Diagram of DLMS4TEFAlgorithm

The full diagram of DLMS4TEF's algorithm is shown in figure (2), this diagram reveals all sub-algorithms and their relationship among each others, so it is very important to understand DLMS4TEF's algorithm.

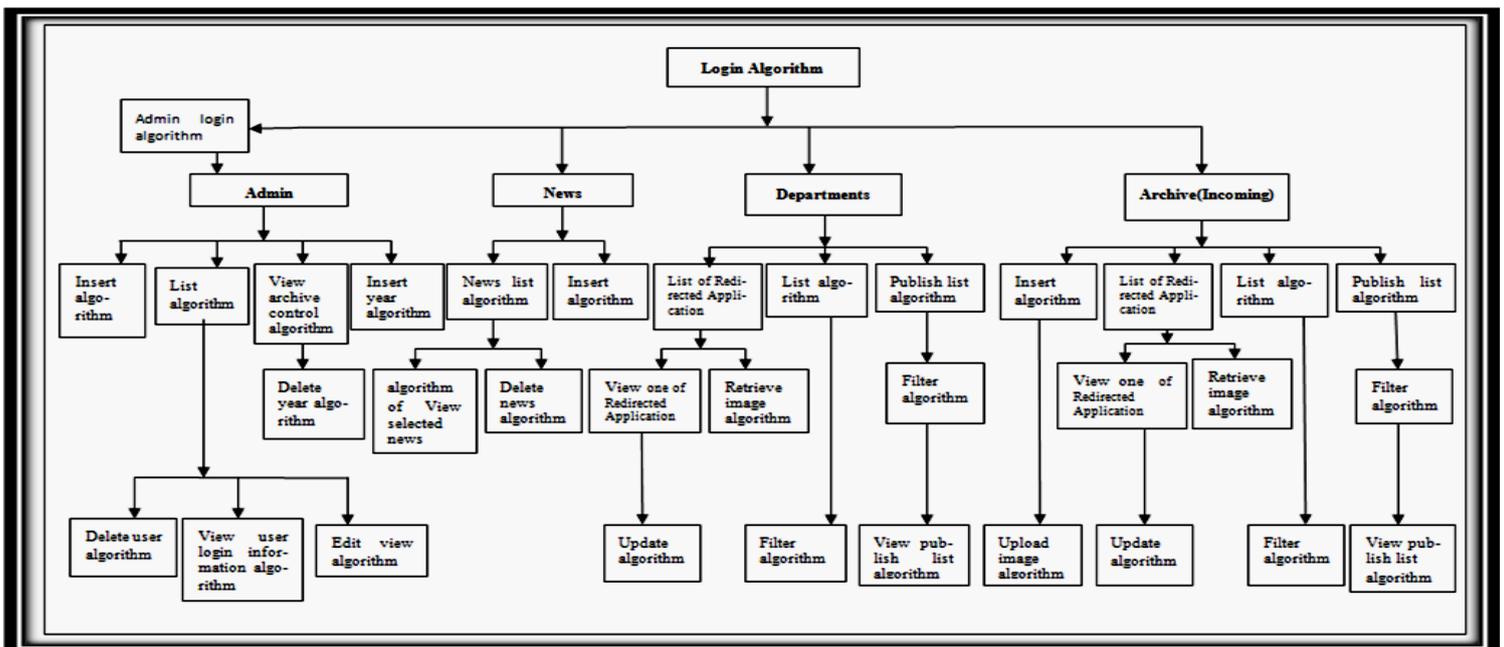

*Figure (2): Diagram of DLMS4TEF's algorithm*

- ❖ Login Algorithm: Represents the main entry to work with DLMS4TEF, so it needs the user-ID which is defined for each user/client in the department previously, and then it will be saved in users table to achieve the login process.
- ❖ List of Redirected Applications Algorithm: For redirecting applications, the user needs to redirect it to correct path/section which has been done by the following algorithm.
- ❖ Retrieve Image Algorithm: Allows retrieving the image name from database, then displays a specific application image.
- ❖ Insert Algorithm: Allows user to enter the application information in the data table.
- ❖ Upload Images: Algorithm uploads some application type's image.
- ❖ List Algorithm: The steps of this algorithm explain how to show the required fields of all applications after inserting them in data base, so its benefit of is monitoring the path of application.
- ❖ Filter Algorithm: It explains the steps that the user needs to reach to the information of the specific application.
- ❖ Update Algorithm: The steps of this algorithm are needed for exchanging or updating application's information before redirect it to another department.

## 2-4 DLMS4TEF of Database

To implement any web application like DLMS4TEF, it is needed to use database for dynamic, i.e. it must base on some databases to do multiple functions/tasks, e.g. storage, retrieve, update, delete, etc, and it can't implement any dynamic application without using database. Thus, the structure of DLMS4TEF's database and its tables has been explained. For DLMS4TEF website, it is used MySQL package to design DLMS4TEF's database. The database is called project

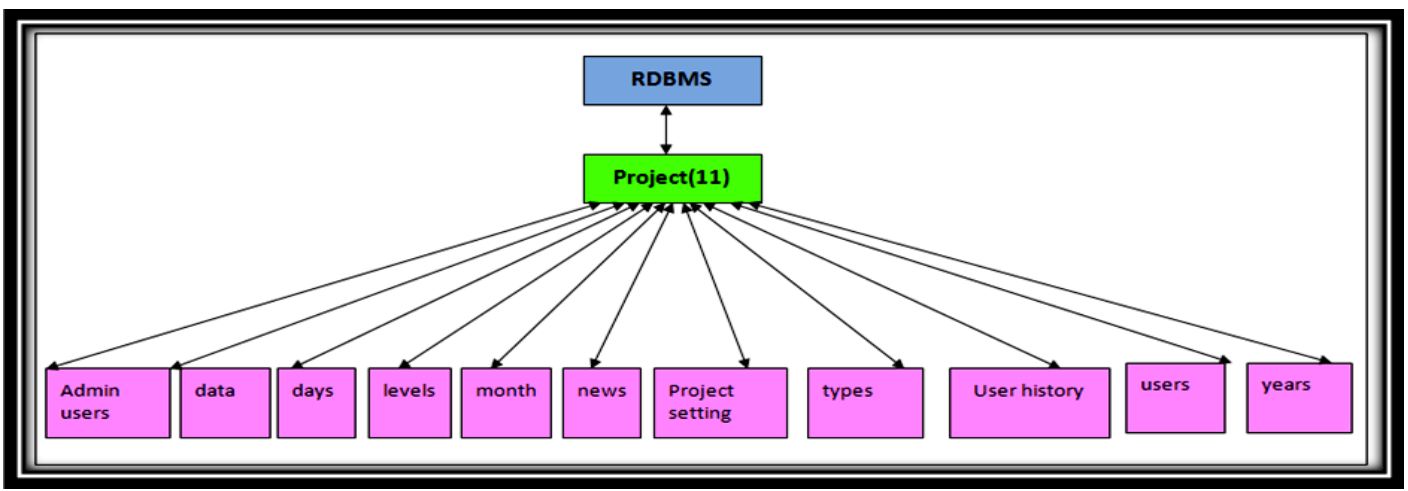

*Figure (3): Diagram of DLMS4TEF's algorithms*

## 3 – The Implementing of DLMS4TEF

To implement a proposal design of DLMS4TEF as local website which is referring to in section 2, and to satisfy all the aims which are referring to in subsection (1-2), will use WAMP5 Server version (1.7.4) which is support PHP to write all scripts which are needed and MySQL to create DLMS4TEF's database (project11) as referring to in subsection (2-4).

implement DLMS4TEF required as first built client/server network with configurations (hardware and software) which are referring to in subsection (2-1), and determine the permissions from administrator for each client in department of foundation's departments for more security using HTAccessible program which allows the user to access by using the IP address, i.e. the server will reject any tasks/functions when comes from unexpected client in department, finally install DLMS4TEF and supported by Kurdish language using Kurdish Unicode to make the client can work with Kurdish GUIs as shown below, so for any application on DLMS4TEF open the browser, then enter **URL** of DLMS4TEF (**www.foundation.edu**) to reach index page as shown in figure (4).

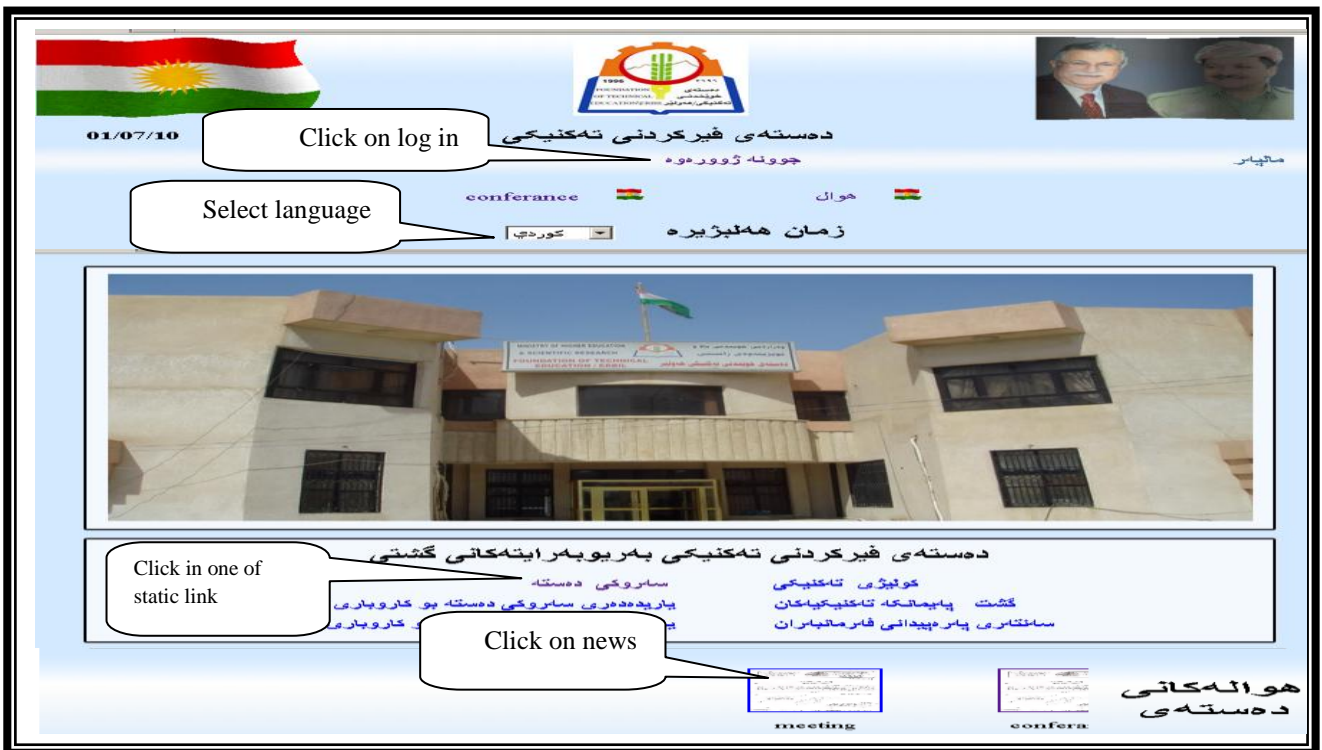

*Figure (4): Index page*

When login hyperlink is clicked on, the login page will be viewed as shown in figure (5).

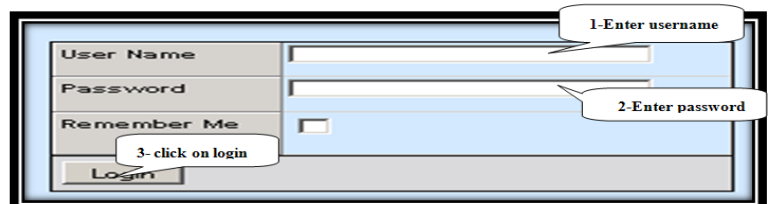

*Figure (5): Login page*

Entering the department's username and password, and then clicking on the login button will reveal the specific department page; when a specific IP address of a user is denied access to another department, this inaccessibility will display message as shown in figure (6).

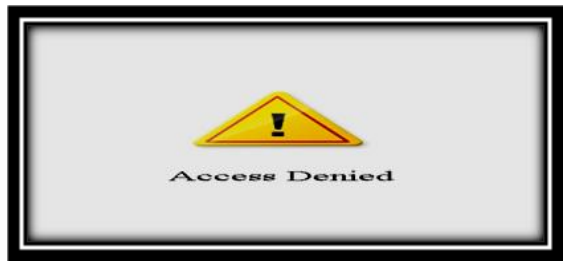

*Figure (6): Access denied message*

Figure (7) shows the home page for incoming-archive (incoming section).

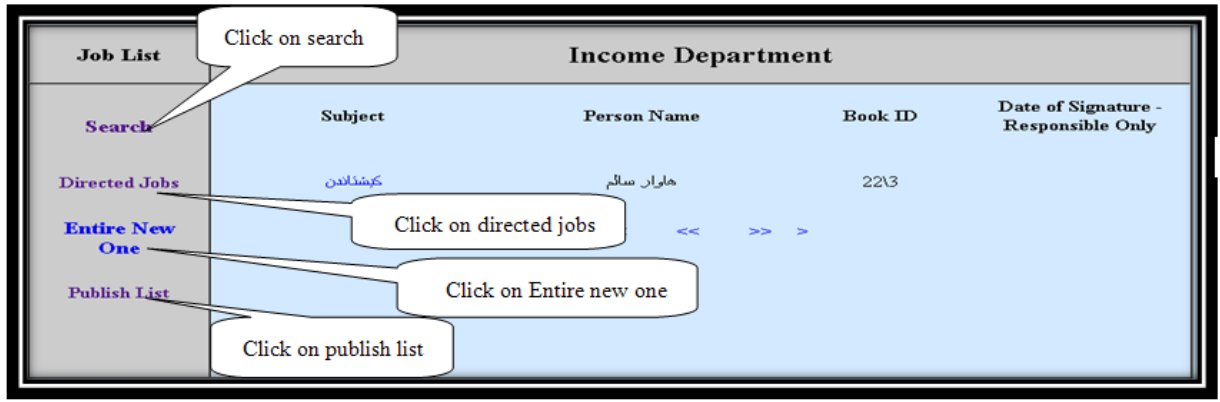

*Figure (7): Incoming index page*

This department includes the following functions:

**A- Search hyperlink:** this helps to view the applications list which is inserted in the database with its required field of information as shown in figure (8).

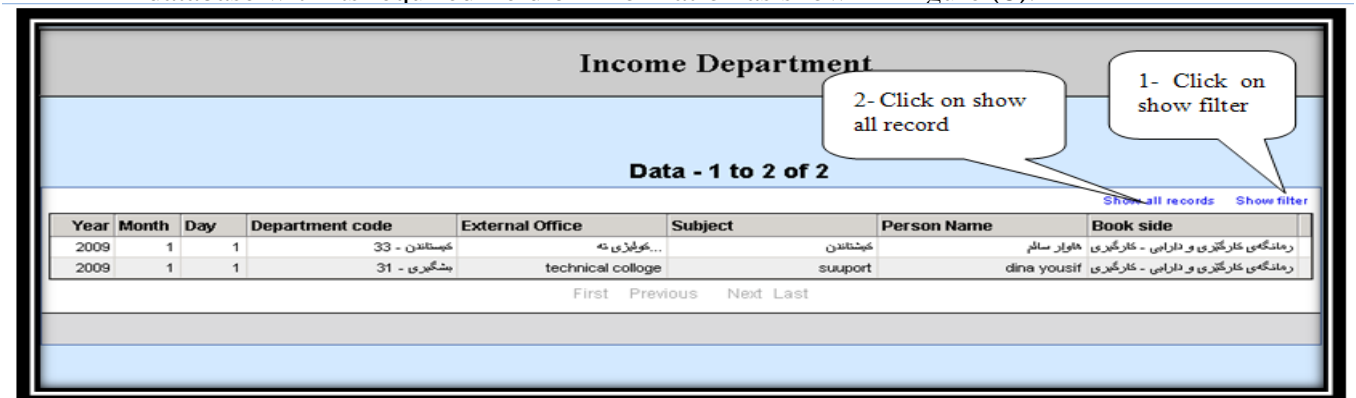

*Figure (8): Search page*

Clicking on the filter button will display the filter page as referred to as shown in figure (9).

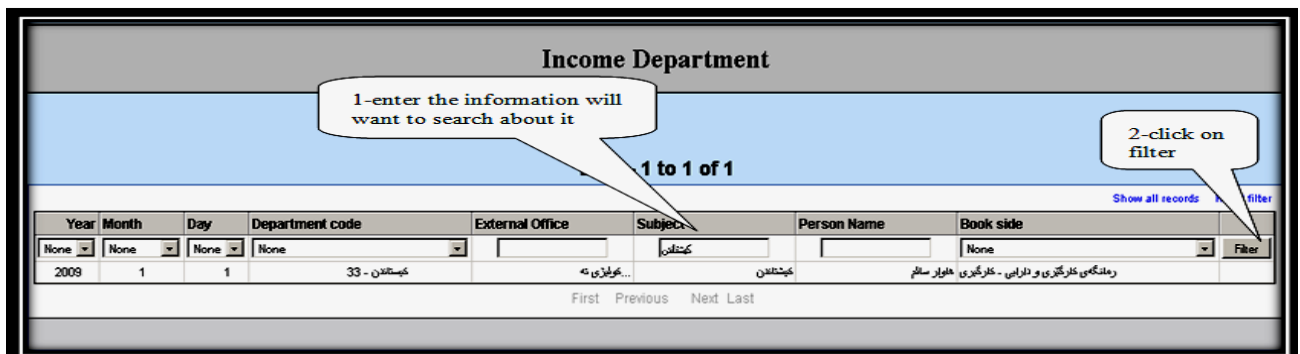

*Figure (9): Filter page*

To display all records after filtering, click on show all records, as shown in figure (9) above.

B- **Directed jobs hyperlink**: to view and display the directed jobs list as shown in figure **(7)**, then click on the specific application to display it.

C- **Entire new one hyperlink:** it is used in viewing the insert page of the incoming section, as shown in figure (10), and working with some applications.

*Figure (10): insert page*

D- **Publish list hyperlink**: this is used in viewing the exported application list as referred to in and enabling the department's client to see these applications after they are outgoing from the outgoing department, as shown in figure (11).

*Figure (11): publish list page*

After exporting the application, click on its subject name to view the application's information.

Through implementing and applying DLMS4TEF, it is noted important features that need to be discussed, whether individually or by comparison with other projects. These important features are as follows:

1. Clarifying the size of the database after doing backup the natural size of the DLMS4TEF database (project) is 172KB before exporting. After exporting, the database (project) has more than one option for compression. When implementing the first one, which is NONE, the database size of DLMS4TEF will decrease from 172KB to 22.4KB, i.e. the size will reduce to ~13% from the natural size, but when using the zipped option for compression, the project size will decrease from 172KB to 5.22KB, i.e. the size will reduce to ~3%
2. Providing comparisons between the traditional management in the Technical Education Foundation-Erbil , the proposal design (DLMS4TEF), and one of the projects that are currently underway in some neighboring countries, such as Oman's project, Table (1) presents comparisons between the traditional, DLMS4TEF in the Kurdistan Region and Oman's project.

*Table (1): comparisons between* DLMS4TEF*, traditional and Oman project.*

| Feature | Traditional | Oman project [7] | DLMS4TEF |
|---|---|---|---|
| **Benefits of Networks.** | Haven't | Have | Have |
| **LAN Technology** | Hasn't LAN technology. | Using FDDI and ATM technologies | Using Ethernet as LAN technology, make it more efficient. |
| **Speed and High Performance** | No | Yes | Yes maybe more than Oman project, if used Giga-Ethernet in Future. |
| **Deployment Cost (relative to each other)** | Cheap | Expensive | Cheap, comparing with Oman, because doesn't use FDDI or ATM technology |
| **Monitoring and Security** | limited | Yes | Yes |
| **Apply new software** | No | No | Yes, like apply Kurdish Unicode, which is used for Kurdish GUI, so it makes DLMS4TEF easy in use. |

## 4- Conclusions

Through the implementation and application of DLMS4TEF, the following conclusions are reached:

1. DLMS4TEF has been supplied Kurdish GUIs; therefore, it is easy and general to use it in the Kurdistan Region.
2. DLMS4TEF takes part in connecting the departments and directorates of the Technical Education Foundation all together with Ethernet (LAN) technology and client-server network. Moreover, it will provide all benefits of network, such as sharing resources, low cost, security, etc.

3. DLMS4TEF has been contributed to the automatic management across local dynamic websites of the Technical Education Foundation. This feature allows the staff not to leave their positions or waits for the person who is in charge of delivering the applications files to other departments, because the movement of the application files becomes through that local website. This will reduce the contacts between the citizens and departments; i.e. the contacts will be limited to those between the citizens and the reception or the archive, because one of the DLMS4TEF functions is tracking the application.

4. There is the possibility of adding, view/display, and deleting news. This feature will be used in notification/memorize commands spontaneously to the employees of the foundation.

5. In the admin. Department, it provides the possibility of making backup to DLMS4TEF's databases at the end of each day and note the efficient technique to minimize the size of the database in using the option (zipped) which reaches ~3% of the original size of the database.

# PolyTechnic

## CONTENTS